\documentclass{elsart}
\usepackage{graphics,natbib,graphicx,psfig,amssymb,ccaption} 
\usepackage{lscape} 
\journal{New Astronomy}
\input psfig.sty
\begin{document}
\begin{frontmatter}
\title{Population types of cataclysmic variables in the solar 
neighbourhood}
\author[istanbul]{T. Ak\corauthref{cor}},
\corauth[cor]{corresponding author.}
\ead{tanselak@istanbul.edu.tr}
\author[istanbul]{S. Bilir},
\author[sabanci]{T. G\"uver},
\author[istanbul]{H. \c Cakmak},
\author[istanbul]{S. Ak}
\address[istanbul]{\. Istanbul University, Faculty of Sciences, 
Department 
of Astronomy and Space Sciences, 34119 University, \. Istanbul, 
Turkey}
\address[sabanci]{Sabanc\i  ~University, Faculty of Engineering 
and Natural 
Sciences, Orhanl\i --Tuzla, 34956 \. Istanbul, Turkey}

\begin{abstract}

The Galactic orbital parameters of 159 cataclysmic variables in the 
Solar neighbourhood are calculated, for the first time, to determine 
their population types using published kinematical parameters. 
Population analysis shows that about 6 per cent of cataclysmic variables 
in the sample are members  of the thick disc component of the Galaxy. 
This value is consistent with the fraction obtained from star count 
analysis. The rest of the systems in the sample are found to be in the 
thin disc component of the Galaxy. Our analysis revealed no halo CVs in 
the Solar vicinity. About 60 per cent of the thick disc CVs have orbital 
periods below the orbital period gap. This result is roughly consistent 
with the predictions of population synthesis models developed for 
cataclysmic variables. A kinematical age of 13 Gyr is obtained using total 
space velocity dispersion of the most probable thick disc CVs which is 
consistent with the age of thick disc component of the Galaxy.

\end{abstract}

\end{frontmatter}

\section{Introduction}

Cataclysmic variables (CVs) are short-period interacting binary stars, 
consisting of a white dwarf primary and a low-mass late spectral type 
secondary star. The secondary star fills its Roche lobe and transfers 
matter to the primary via a gas stream and an accretion disc. Accretion 
disc formation is prevented in systems with strongly magnetic white 
dwarfs in which mass accretion continues through accretion columns. For 
a detailed review of observational properties of CVs, see \citet{War95}.

\citet{SKR97} studied the long-term evolution of cataclysmic variables 
as a function of the secondary star metallicity. They showed that Pop 
II CVs with a low metallicity secondary star have a detached phase with 
a smaller orbital period width, a shorter minimum period \citep{Pac67} 
and a slightly 
higher mass transfer rate, resulting in shorter evolutionary timescales 
compared to CVs where the secondary star has a Solar chemical 
composition. According to their population synthesis model, most Pop II 
CVs are expected to be found below the period gap 
\citep{VZ81,Retal82,Retal83,PS83,SR83,K88,SL12}. \citet{SKR97} express 
that the high $\gamma$ velocities (systemic velocities or center of mass 
velocities) of some systems found by \citet{vPaetal96} suggest these 
systems to be Pop II CVs. However, most of these CVs are magnetic systems 
(DQ Her and AM Her stars), where the Doppler-shifts of spectral lines 
originate mainly from the accretion stream. Consequently, the errors in 
the $\gamma$ velocities may be noticeably high. Still, finding magnetic 
systems below the period gap should not be a surprise as these systems 
concentrate towards shorter orbital periods, with little evidence for 
a period gap \citep{War95}. Interestingly, this narrow (or none) period 
gap is consistent with the predictions from the study of \citet{SKR97}. 
It should be noted that \citet{Aketal10} found magnetic systems to be 
much older than non-magnetic systems while they also emphasized doubts 
about the reliability of $\gamma$ velocities obtained from the 
observations of magnetic systems. 

Some CVs have been suggested to be the members of old populations in the 
Galaxy \citep{HW87,HMW87,HS90,Drissenetal94,Sheetsetal07,Uthasetal11,IT12}. 
Clearly, it is hard to detect thick disc and halo CVs if they are not 
outbursting systems. CVs in globular clusters could be detected from their 
outbursts, emission lines, X-rays or their blue colours. Low metallicity 
values extracted from spectra can be another indicator for thick disc and 
halo CVs \citep{SKR97}. However, spectroscopic confirmation of these 
detections is difficult as they are very faint systems. For a detailed 
discussion on the detection of CVs in globular clusters, we refer to 
Knigge's (\citeyear{Kng11}) review and references therein. 

A reasonable observable sample of Pop II CVs (thick disc and halo CVs) 
can only be found at vertical distances from the Galactic plane 
$z$ $\gtrsim$ 2 kpc \citep{SKR97}. 
However, the faintness of CVs restricts most of their photometric and 
spectroscopic studies to the Solar neighbourhood. Although some abundance 
anomalies in UV and IR were found for CVs 
\citep{Hamiltonetal2011,Gansickeetal2005}, there are no reliable metallicity 
measurements for CVs. In the absence of metallicity measurements, only 
kinematical and the dynamical methods are expected to provide reliable 
results in the recognition of the thick disc and halo CVs in the Solar 
neighbourhood. Although the development of sensitive 
CCD cameras, spectrographs and larger size telescopes made it possible to 
observe relatively faint CVs, we lack the necessary observational 
information to understand the kinematical properties of CVs in the Galactic 
scales.  

In the next section we use the kinematical properties of 159 CVs in the solar 
neighbourhood to distinguish different Galactic populations among them.

\section{The data}

\subsection{Input Parameters}

The kinematical data used in this study are taken from \citet{Aketal10}. The 
most important inputs used in their study are distances, $\gamma$ velocities 
and proper motions. The proper motions of CVs were mostly obtained from the 
NOMAD Catalogue \citep{Zachetal05}. The types, equatorial coordinates and 
orbital periods of CVs were mostly taken from \citet[][Edition 7.7]{RK03} 
and \cite{Downetal01}.

The distances were predicted using the PLCs relation of \citet{Aketal07}. The 
PLCs relation is based on orbital period and Two Micron All Sky Survey \cite[2MASS,][]{Strutskie06} 
JHK$_s$ photometric data \citep{Cutri03}. This relation is reliable and valid in the 
ranges $0.032$ $< $P$(d) \leq 0.454$, 
$-0.08 < (J-H)_{0} \leq 1.54$, $-0.03 < (H-K_{s})_{0} \leq 0.56$ and 
$2.0 < M_{J} < 11.7$ mag. For a detailed description of the method by the 
PLCs relation, we refer to Ak et al.'s (\citeyear{Aketal07,Aketal08}) studies . 
The distances obtained from this relation differ in general less than 4$\%$ 
from those obtained from trigonometric parallaxes \citep{GarRing12}.

The last important inputs used in \citet{Aketal10} are the radial velocities 
which are used to calculate total space velocities. The radial velocity with 
respect to the Sun comes from measurements of Doppler-shifts in spectral 
lines. However, one has to take here into account that CVs are binaries, 
and thus one has to determine the radial velocity of the center of mass of 
the system. \citet{Aketal10} adopted the criteria defined by \citet{vPaetal96} 
when collecting 
$\gamma$  velocities from the literature and merged these $\gamma$ velocities 
with those collected by \citet{vPaetal96}. Radial velocities, and 
consequently $\gamma$  velocities, derived from emission lines are likely 
affected by the motion in the accretion disc or the matter stream falling 
on the disc from the secondary. Thus, \citet{Aketal10} analyzed $\gamma$ 
velocities statistically and looked for possible systematic errors in the 
$\gamma$ values obtained from emission lines. They concluded that there 
is no substantial systematic difference between systemic velocities derived 
from emission and absorption lines and that the observed $\gamma$ velocities 
can be reliably used for statistical analysis \citep[see][for details]{Aketal10}.   

From the celestial coordinates ($\alpha$, $\delta$), proper motion 
components ($\mu_\alpha\cos\delta$, $\mu_\delta$), systemic velocity 
($\gamma$) and the parallax ($\pi$), \citet{Aketal10} computed Galactic 
space velocities and their propagated errors with respect to the Sun using 
the algorithms and transformation matrices of \cite{JS87}. Although the sampled 
CVs are relatively nearby objects, \citet{Aketal10} applied corrections for 
differential 
Galactic rotation to space velocities as described in \citet{MB81}. 
\citet{Aketal10} analyzed the propagated errors of space velocities, with 
respect to LSR (Local Standart of Rest), and refined their sample by 
removing systems with a total space velocity error $S_{err}$ $>$ 30 km s$^{-1}$. 
Although they collected input data from the literature for 194 CVs, this 
analyses decreased the number of usable systems in their sample to 159. In this 
study, the final sample of 159 CVs in \citet{Aketal10} is used to calculate 
the Galactic orbital parameters of systems. 

\subsection{Calculation of Galactic Orbits}

In order to determine possible Galactic orbits of CVs, we first perform 
test-particle integration in a Milky Way potential which consists of 
a logarithmic halo of the form 

\begin{eqnarray}
  \Phi_{\rm halo}(r)=v_{0}^{2} \ln \left(1+\frac{r^2}{d^2}\right),
\end{eqnarray}
with $v_{0}=186$ km s$^{-1}$ and $d=12$ kpc. The disc is represented by 
a Miyamoto-Nagai potential:

\begin{eqnarray}
  \Phi_{\rm disc}(R,z)=-\frac{G M_{\rm d}} { \sqrt{R^{2} + \left(
        a_d + \sqrt{z^{2}+b_d^{2}} \right)^{2}}},
\end{eqnarray}
with $M_{\rm d}=10^{11}~M_{\odot}$, $a_d=6.5$ kpc and $b_d=0.26$
kpc. Finally, the bulge is modelled as a Hernquist potential,

\begin{eqnarray}
  \Phi_{\rm bulge}(r)=-\frac{G M_{\rm b}} {r+c},
\end{eqnarray}
using $M_{\rm b}=3.4\times10^{10}~M_{\odot}$ and $c=0.7$ kpc.  The 
superposition of these components gives a good representation of the Milky 
Way. The circular speed at the Solar radius is taken $\sim$220 km s$^{-1}$. 
The orbital period of the LSR is $P_{LSR}=2.18\times10^8$ years while 
$V_c=222.5$ km s$^{-1}$ denotes the circular rotational velocity at the 
Solar Galactocentric distance, $R_0=8$ kpc. The same formulae were already 
used to determine the Galactic orbits of objects from different classes, 
e.g. \cite{Coskunogluetal12} and \cite{Bil12}. 

In order to analyse Galactic orbits of CVs, the mean radial Galactocentric 
distance ($R_m$) is taken into account as a function of the stellar 
population and the orbital shape. We consider the planar and vertical orbital 
eccentricities, $e_p$ and $e_v$, respectively. $R_m$ is defined as the 
arithmetic mean of the final perigalactic ($R_p$) and apogalactic ($R_a$) 
distances, and $z_{max}$ and $z_{min}$ are the final maximum and minimum 
distances, respectively, to the Galactic plane whereas $e_p$ and $e_v$ are 
defined as follows:

\begin{eqnarray}
e_p=\frac{R_{a}-R_{p}}{R_{a}+R_{p}},
\end{eqnarray}

and

\begin{eqnarray}
e_v=\frac{|z_{max}|+|z_{min}|}{R_m},
\end{eqnarray}
respectively, where $R_m=(R_a+R_p)/2$ \citep{Vidojevic09}. Due to 
$z$-excursions $R_p$ and $R_a$ can vary, however this variation is not 
more than 5$\%$. Calculated orbital parameters of the 159 CVs are listed 
in Table 1. The columns of the table are organized as follows: name, equatorial 
($\alpha, \delta$) coordinates, the final perigalactic ($R_p$) and 
apogalactic ($R_a$) distances, the maximum ($z_{max}$) and minimum 
($z_{min}$) distances to the Galactic plane, total orbital angular momentum 
($J_{z}$), and the planar ($e_p$) and vertical ($e_v$) orbital 
eccentricities. Apogalactic and perigalactic distances are determined from 
the averaged maximum and minimum galactocentric distances of systems in the 
calculated Galactic orbits within the integration time of 3 Gyr, i.e. backwards 
in time over an interval of 3 Gyr. This integration time is chosen to 
correspond to 12 or 15 revolutions around the Galactic center so that the 
averaged orbital parameters can be determined reliably.

\begin{table*}
\setlength{\tabcolsep}{4.2pt}
{\scriptsize
\begin{center}
\caption{The data sample. The table can be obtained electronically.}
\begin{tabular}{rlccccccccc}
\hline
 ID & Star & $\alpha$~(J2000.0) & $\delta$~(J2000.0) & $R_p$ & $R_a$ &  $z_{min}$  
 &  $z_{max}$ & $J_{z}$  &  $e_v$ & $e_p$  \\
 &  & (hh:mm:ss) & ($~^{\circ}$~:~${'}$~:~${''}$~) & (kpc) & (kpc) & (kpc) 
 & (kpc) & (kpc km s$^{-1}$) &  &  \\
\hline
  1 & AR And   &   01:45:03.28   &   +37:56:32.7               
  & 7.894 & 8.207 &   -0.124    &    0.124   & 1790.1 &  0.015 & 0.019   \\
  2 & LX And   &   02:19:44.10   &   +40:27:22.9               
  & 6.517 & 9.010 &   -0.150    &    0.150   & 1681.0 &  0.019 & 0.160   \\
  3 & PX And   &   00:30:05.81   &   +26:17:26.5               
  & 7.001 & 9.804 &   -0.492    &    0.491   & 1815.7 &  0.058 & 0.167   \\
  4 & RX And   &   01:04:35.54   &   +41:17:57.8               
  & 7.984 & 8.525 &   -0.128    &    0.128   & 1837.4 &  0.016 & 0.033   \\
  5 & V603 Aql &   18:48:54.64   &   +00:35:02.9               
  & 7.441 & 7.957 &   -0.081    &    0.081   & 1705.9 &  0.010 & 0.034   \\
  . &  .....   &     ......      &     ......                  
  & ....  & ....  &   .....     &    ....    &  ..... &  ....  & ....    \\
  . &  .....   &     ......      &     ......                  
  & ....  & ....  &   .....     &    ....    &  ..... &  ....  & ....    \\
  . &  .....   &     ......      &     ......                  & ....  
  & ....  &   .....     &    ....    &  ..... &  ....  & ....    \\
\hline
\end{tabular}
\end{center}
}
\end{table*}

\subsection{Determination of Population Types}

In order to determine the population types of the CVs, we define some dynamical 
and kinematical criteria. The criteria defined to select thick disc 
or halo CVs are as following: 
\begin{enumerate}
 \item[(1)] The total space velocity ($V_{tot}$). \cite{Afsaretal12} 
 showed that stars with $V_{tot}\geq$ 100 km s$^{-1}$ are thick disc or 
 halo objects. 
  \item[(2)] The relative probability for thick disc to thin disc 
  membership ($TD/D$). \cite{Bensetal03,Bensetal05} proposed that the stars 
  with $TD/D > 1$ are thick disc or halo objects.
 \item[(3)] The maximum distance from the Galactic plane ($z_{max}$). It 
 is well known that the thick disk is the dominant component of the 
 Galaxy between 1 and 5 kpc above the Galactic disc while halo 
 objects have $z_{max}$ values larger than 5 kpc \citep{Bil08}.
  \item[(4)] Vertical orbital eccentricity ($e_v$). \cite{Bil12} concluded 
 that vertical eccentricities of Galactic 
 orbits calculated for the thick-disc and halo stars are larger than $\sim$0.1. 
 \item[(5)] Planar eccentricity of the Galactic orbit ($e_p$). Using 
 main-sequence stars, \cite{Paulietal2003} found that stars with 
 $e_p$ $\gtrsim$ 0.3 belong to the thick disc. 
\end{enumerate}

\begin{table*}
\setlength{\tabcolsep}{5.4pt}
\scriptsize{
\begin{center}
\caption{Probable thick disc CVs in the sample. Remarks indicate 
the fulfilled criteria as defined in the text. }
\begin{tabular}{llccccccl}
\hline
 & & & \multicolumn{5}{c}{Criteria} & \\
 & & & (1) & (2) & (3) & (4) & (5) &\\
\hline
  Star         & Type     & $P_{orb}$ &  $V_{tot}$    & $TD/D$    
  & $z_{max}$ & $e_v$ & $e_p$ & Remark \\
               &          &    (hr)   & (km s$^{-1}$) &           
               &   (kpc)   &       &       &  \\
\hline

  J2050-0536   & NL, AM   &   2.299   &   100.55 &  14.5  & 0.46  
  & 0.038 & 0.381 & 1, 2, 5 \\
  VV Pup       & NL, AM   &   1.674   &   151.63 &  9602  & 0.54  
  & 0.036 & 0.511 & 1, 2, 5 \\
  BY Cam       & NL, AM   &   3.354   &   142.74 &  1064  & 0.80  
  & 0.060 & 0.456 & 1, 2, 5 \\
  CT Ser       & N        &   4.680   &   80.58  &  0.2   & 1.00  
  & 0.145 & 0.243 & 3, 4 \\
  V1043 Cen    & NL, AM   &   4.190   &   80.59  &  8.3   & 1.03  
  & 0.102 & 0.229 & 2, 3, 4    \\
  J1629+2635   & NL       &   2.033   &   72.11  &  2.5   & 1.04  
  & 0.112 & 0.184 & 2, 3, 4    \\
  IY UMa       & DN, SU   &   1.774   &   114.73 &  70    & 1.39  
  & 0.148 & 0.284 & 1, 2, 3, 4 \\
  J0813+4528   & DN, UG   &   6.936   &   78.34  &  0.5   & 1.77  
  & 0.189 & 0.198 & 3, 4       \\
  AK Cnc       & DN, SU   &   1.562   &   154.72 &  36015 & 2.26  
  & 0.195 & 0.413 & 1, 2, 3, 4, 5 \\
\hline
\end{tabular}
\\
\end{center}
{\tiny DN: dwarf novae, NL: nova-like stars, N: novae, SU: SU UMa type 
dwarf novae, UG: U Gem type dwarf novae, AM: AM Her 
(polars) type nova-like stars.}
\\
}
\end{table*}

$TD/D$ is calculated using a purely kinematical approach described by 
\cite{Bensetal03}. These probabilities were estimated and listed by 
\cite{Aketal10}. According to the criteria defined by 
\cite{Bensetal03} the stars are selected from four different $TD/D$ intervals: 
$TD/D < 0.1$ (i.e. ``high probability thin-disc stars''); $0.1< TD/D < 1$ 
(i.e. ``low probability thin-disc stars''); $1< TD/D < 10$ (i.e. ``low 
probability thick-disc stars'') and $TD/D > 10$ (i.e. ``high probability 
thick-disc stars''). So, in this study a possible thick disc or halo CV is 
expected to fullfill at least two of these criteria.

Using the criteria defined above, we find that nine of 159 CVs in the 
sample belong to thick disc population. The rest of the sample consists of 
thin disc systems. It is concluded that there are no halo CVs, as the maximum 
distances to the Galactic plane $z_{max}$ of the CVs in the sample do not 
exceed 5 kpc. Thick disc CVs found in this study are 
listed in Table 2 with their total space velocities ($V_{tot}$), with respect 
to LSR, types, orbital periods ($P_{orb}$), and the maximum distances to the 
Galactic plane ($z_{max}$). The relative probability for the thick disc to 
thin disc membership ($TD/D$) is also listed in the Table 2. $e_p$ and $e_v$ 
are planar and vertical eccentricities of the Galactic orbits, respectively. 
In Figs. 1-3, nine thick disc CVs are indicated in 
$V_{tot}$-$e_p$, $e_p$-$e_v$ and $e_p$-$J_z$ diagrams of the CV sample in 
this study, respectively.


\begin{figure}
\begin{center}
\includegraphics[scale=0.4, angle=0]{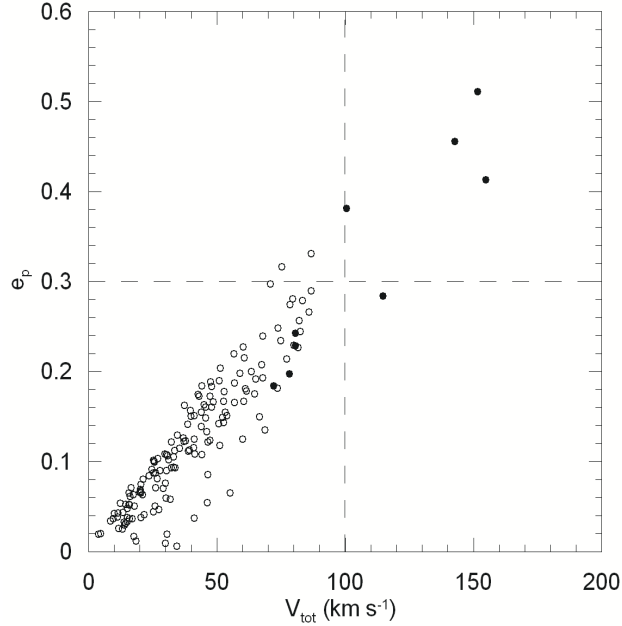}
\caption[] {\small $V_{tot}$-$e_p$ diagram for the CVs in the sample. 
Thick disc and thin disc objects are indicated with filled 
and empty circles, respectively. Dashed lines represent limit values 
of criteria defined in the text.}
\end{center}
\end{figure}


\begin{figure}
\begin{center}
\includegraphics[scale=0.4, angle=0]{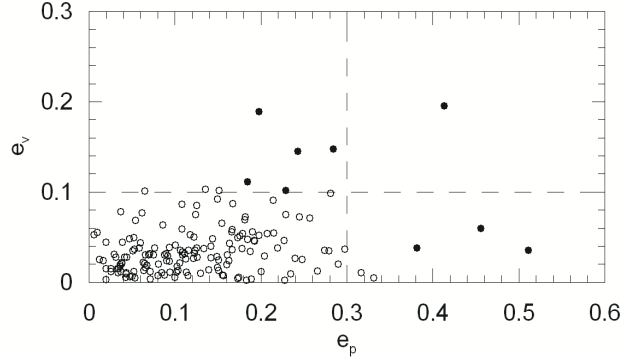}
\caption[] {\small $e_p$-$e_v$ diagram for the CVs in the sample.
Symbols and dashed lines are as in Fig. 1}
\end{center}
\end{figure}


\begin{figure}
\begin{center}
\includegraphics[scale=0.4, angle=0]{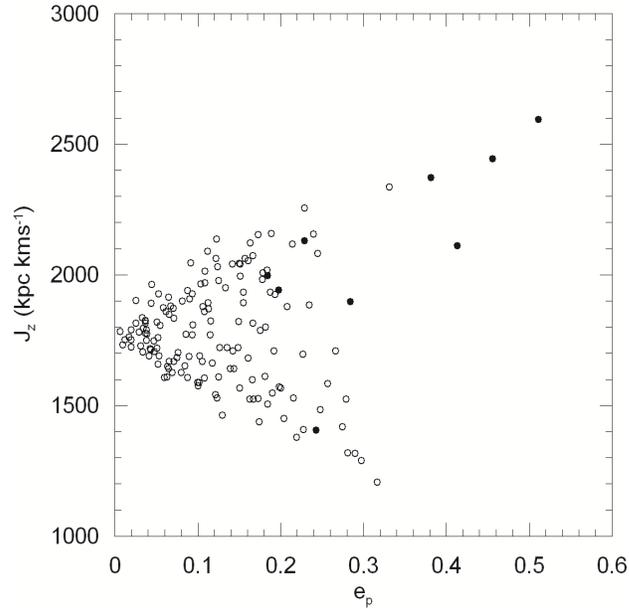}
\caption[] {\small $e_p$-$J_z$ diagram for the CVs in the sample. 
Symbols are as in Fig. 1}
\end{center}
\end{figure}

AK Cnc is the most probable thick disc CV in the sample. This object 
fulfills all the criteria we defined above. IY UMa is also one of the 
strongest thick disc candidates of CVs. IY UMa accomplishes four criteria 
but one: the planar eccentricity of its Galactic orbit is 0.284 which is very 
close to 0.3 (see, Fig. 3).VV Pup, BY Cam and J2050-0536 accomplish three 
criteria: $V_{tot}\geq$ 100 km s$^{-1}$, $TD/D > 1$ and $e_p$ $\gtrsim$ 0.3 
for these CVs (see, Fig. 2). For J1629+2635 and V1043 Cen, we found 
$TD/D > 1$, $z_{max}$ $\gtrsim$ 1 kpc and $e_v$ $\gtrsim$ 0.1. CT Ser and 
J0813+4528 match with two criteria. We found $z_{max}$ $\gtrsim$ 1 kpc and 
$e_v$ $\gtrsim$ 0.1 for these two systems. AK Cnc, IY UMa, VV Pup, BY Cam 
and J2050-0536 were already suggested to be thick disc objects by 
\cite{Aketal10} using a pure kinematical approach. Our results is consistent 
with this suggestion.

\section{Conclusion and discussions}

In this study we calculated, for the first time, the Galactic orbital parameters 
of a large number of CVs. We also determined their population types using these 
calculations and the kinematical parameters as found by \citet{Aketal10}

Our analysis shows that almost all of the CVs in the sample are in the thin 
disc component of the Galaxy. Only nine CVs in the sample are 
thick disc stars. These objects are listed in Table 2. These are the CVs which 
cross the Galactic disc through their Galactic orbits. The Galactic orbits of 
these systems as projected on to $X-Y$ and $X-Z$ 
planes are shown in Fig. 4. Galactic orbits of the thick disc CVs in 
Fig. 4 were calculated for an integration time of 3 Gyr, corresponding 
to 12-15 revolutions around the Galactic center. As the maximum distances 
from the Galactic plane ($z_{max}$) in Table 2 are not larger than 5 kpc, we 
conclude that there are no halo objects in the sample.


\begin{figure}
\begin{center}
\includegraphics[scale=0.65, angle=0]{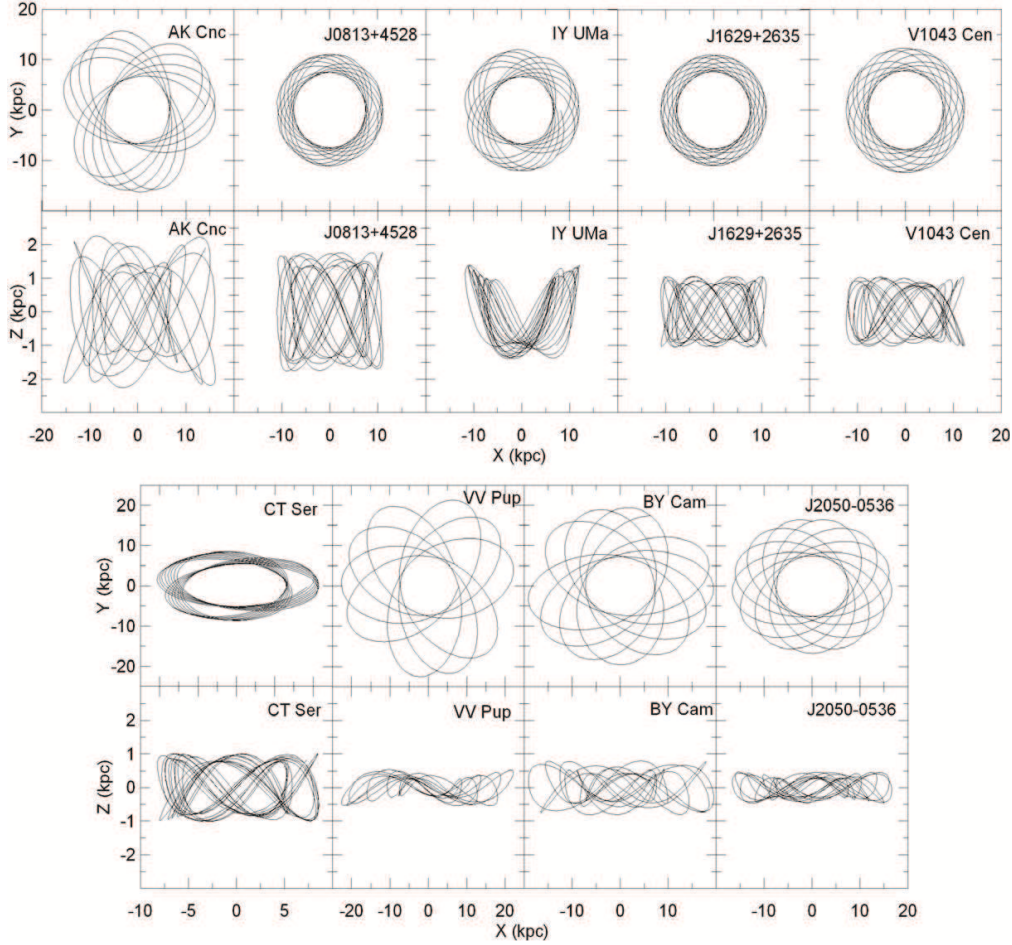}
\caption[] {\small Galactic orbits of the thick disc CVs found in this study 
projected onto $X-Y$ and $X-Z$ planes. Galactic orbits are calculated 
for an integration time of 3 Gyr.}
\end{center}
\end{figure}

It is found that the fraction of thick disc CVs to thin disc CVs in the sample 
is $\sim$6$\%$ which is in very good agreement with the fraction of Pop II 
field stars to Pop I field stars in the Solar neighbourhood 
\citep{Robetal96,Busetal99,Biliretal06}. This suggests that our CV sample is 
complete for the Solar neighbourhood. So, we conclude that statistical studies 
using this sample give reliable and self-consistent results. 

\cite{Aketal10} found that the middle point of the period gap is at 2.62 hr 
for the objects in the sample. An inspection of the thick disc CVs in Table 2 reveals 
that $\sim$60$\%$ of thick disc CVs are located below the orbital 
period gap. According to the population synthesis model of \citet{SKR97}, most 
Pop II CVs are expected to be found below the period gap. We conclude that the 
calculated fraction of thick disc CVs below the period gap to the thick disc 
systems above the period gap in this study is roughly consistent with the 
theoretically expected fraction. It is also interesting to note that 
$\sim$45$\%$ of thick disc CVs found in this study are magnetic systems 
(AM Her stars). 

Some CVs have already been suggested to be the members of old populations in 
the Galaxy. Objects proposed to be Pop II CVs in \cite{HMW87}, 
\cite{Drissenetal94}, \cite{Sheetsetal07}, \cite{Uthasetal11} and \cite{IT12} 
are not included in our sample. \cite{HS90} list 84 known or good candidates 
for being halo CVs by selecting high Galactic latitude objects. Although 21 of 
them are included in our sample, only one of them (AK Cnc) is identified as a 
thick disc CV in this study. It must be emphasized that the Galactic latitude 
is not a reliable indicator for the population type of an object.
Our analysis shows that the most probable thick disc 
CVs in Table 2 have high Galactic angular momentums, implying that they are in 
a population different than the thin disc (Fig. 3). As thick 
disc and halo objects have eccentric orbits, they can pass through the Solar 
neighbourhood during their nuclear evolution. That is why the analysis of Galactic 
orbits is very important in the determination of population types. 

The dispersion of the total space velocities is an indicator of population types, as 
the velocity dispersion of a group of objects is related to their kinematical age.
The total space velocities of the six systems with $z_{max}$ $\gtrsim$ 1 kpc in Table 2 
were taken from \cite{Aketal10} and the dispersion of the total space velocities is 
calculated as 101 km s$^{-1}$. Using the equation given by \cite{Wielen77} and 
\cite{Wielenetal1992}, the kinematical age of the six most probable thick disc 
CVs in the sample is calculated as 13 Gyr. This value is consistent with the age of 
the thick disc population in the Galaxy \citep{FW08}. 

Our study shows that there are thick disc CVs in the Solar neighbourhood. 
Our study also implies that halo CVs, if they are present, must be very rare 
in the vicinity of the Sun. Kinematical studies can reliably prove their presence. 
However, current $\gamma$ velocity measurements are mostly dubious. 
That is why we emphasize the importance of radial velocity studies of 
CVs. Further observational data can help to find more thick disc or halo CVs. 
Specially, investigations based on the data obtained from deep sky surveys 
could help to find more Pop II CVs. Detailed observations of these systems 
could give clues for their evolution.

\section{Acknowledgments}

Part of this work was supported by the Research Fund of the University 
of Istanbul, Project Number: BYP-20366. We thank the anonymous referee 
for his/her comments and suggestions. 
This research has made use of the SIMBAD database, operated at 
CDS, Strasbourg, France. This publication makes use of data products 
from the Two Micron All Sky Survey, which is a joint project of the 
University of Massachusetts and the Infrared Processing and Analysis 
Center/California Institute of Technology, funded by the National 
Aeronautics and Space Administration and the National Science 
Foundation. This research has made use of the NASA/IPAC Extragalactic 
Database (NED) which is operated by the Jet Propulsion Laboratory, 
California Institute of Technology, under contract with the National 
Aeronautics and Space Administration.

\end{document}